# Response of Photon Dosimeters and Survey Instruments to New Operational Quantities Proposed by ICRU RC26

**Thomas Otto**[a]

[a] *Technology Department, CERN,
1211 Genève 23, Switzerland*
E-mail: thomas.otto@cern.ch

ABSTRACT: Operational quantities in radiation protection are defined as conservative estimates of the protection quantities, which are generally not accessible to measurement. Radiation protection dosimeters and monitors are calibrated in the operational quantities. An ICRU Report Committee works on a proposal to replace the operational quantities for external radiation ambient dose equivalent, $H^*(10)$ and personal dose equivalent $H_p(d, \alpha)$ by quantities which are defined in close relation to the protection quantities. In this note, the response of dosimeters and survey instruments for photon radiation to these new operational quantities is evaluated. Survey instruments, having a low-energy cut-off at 50 keV can simply be recalibrated in the new quantity ambient dose and function as usual. Personal dosimeters will show an overreponse to photons with energy below 50 keV and either the dosimeter holder or the evaluation algorithm must be adapted to the new quantity.

.

**Contents**

**1. Introduction**

In protection against ionising radiation, dose limits and optimisation targets are expressed in terms of *protection quantities*. For stochastic effects caused by low doses, the protection quantity is effective dose $E$, defined as the sum of absorbed doses in 15 organs and tissues, weighted by radiation and tissue weighting factors $w_R$, $w_T$:

$$E = \sum_R \sum_T w_T w_R D_{T,R} \quad (1)$$

This quantity is defined over the volume of a human body and is not directly measurable. Effective dose is evaluated by radiation transport calculation with mathematical phantoms of the human body as a target. The latest conversion coefficients between fluence or kerma and effective dose are published in ICRP report 116 [1].

To estimate radiation dose by measurement or calculation, *operational quantities* are required. They are defined in a point in the radiation field and dosimeters and survey instruments are calibrated with respect to them. The presently recommended operational quantities are based on dose equivalent defined in specified depths $d$, for directional dose equivalent $H'(d, \alpha)$ and ambient dose equivalent $H^*(10)$ in the ICRU sphere and for personal dose equivalent $H_p(d, \alpha)$ in the human body under the location of a personal dosimeter. Conversion coefficients to the operational quantities are evaluated by radiation transport calculation in the ICRU sphere or in slab- and cylinder phantoms, the latest numerical values have been published jointly in ICRP Report 74 [2] and ICRU Report 57 [3].

The operational quantities defined in this way have several shortcomings, which can be related to their definition and to details of the numerical calculations. For photons (Figure 1), the published operational quantities $H_p(10)$ and $H^*(10)$ overestimate effective dose at low energies ($E_p < 40$ keV) by up to a factor of five due to the fixed depth of $d = 10$ mm in the phantom, and at high energies ($E_p > 3$ MeV) due to the use of the kerma approximation for their calculation. However, if transport of secondary electrons had been considered in the calculation, the operational quantities would have underestimated effective dose.

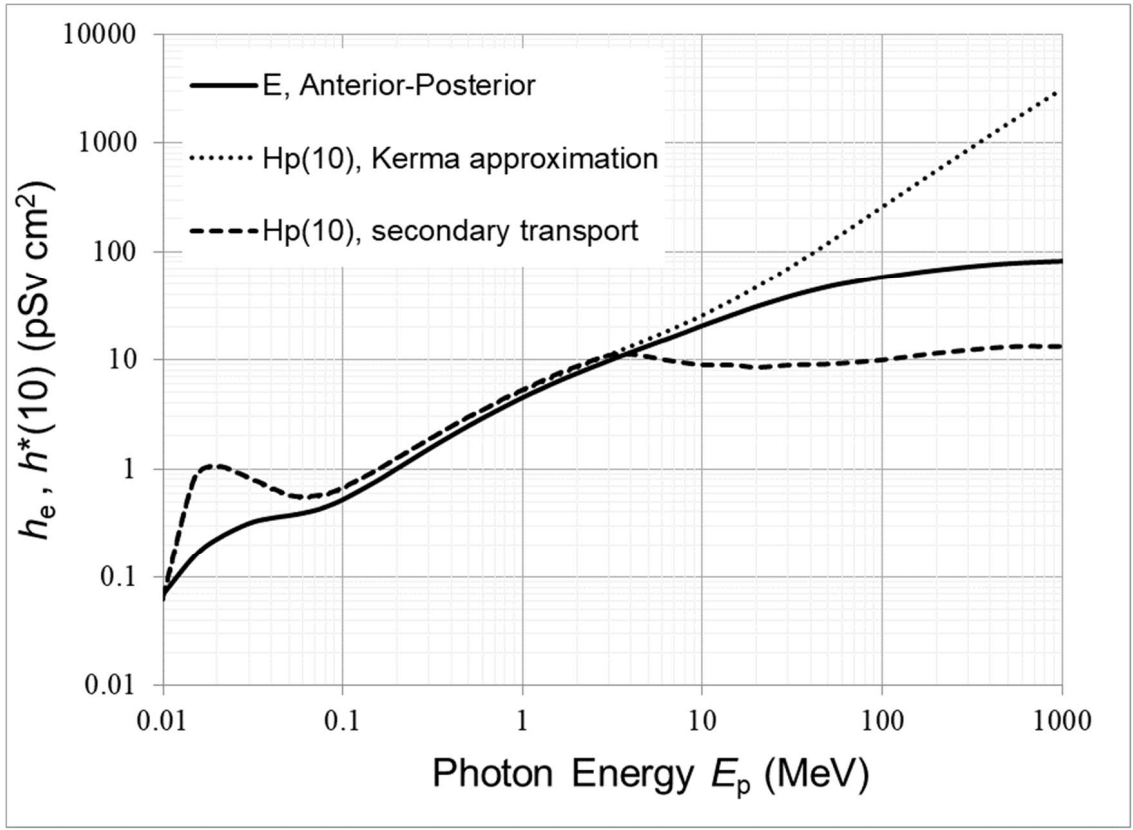

**Figure 1.** Conversion coefficients from photon fluence to a.) effective dose in AP orientation $E(AP)$ (continuous line), b.) personal dose equivalent $H_p(10)$ as published, calculated in kerma-approximation (dotted line) and c.) personal dose equivalent $H_p(10)$ calculated with full electron transport (dashed line).

The ICRU has tasked Report Committee 26 to propose a redefinition of operational quantities [4], [5]. In its report, the committee proposes to introduce operational quantities directly based on the corresponding protection quantity. *Ambient dose $H^*$* is defined for a specified particle type and energy as the maximum value of effective dose $E$ for the different orientations of the radiation field (AP, LLAT, RLAT, PA, ROT, ISO). *Personal dose $H_p(\alpha)$* is calculated in the same anthropomorphic phantom as $E$ for a set of angles of incidence with respect to the vertical axis of the phantom. In particular, $H_p(0^0)$ is numerically equivalent to $E$ (AP).

For the monitoring of deterministic detriment to the skin, the operational quantity *absorbed dose to local skin $D_{local\,skin}(\alpha)$* is introduced, it is calculated in a slab phantom in a depth between 50 and 100 μm, the location of the sensitive layer of skin. Figure 2 shows the comparison between conversion coefficients from kerma in air $K_a$ to the previous and the new operational quantities for personal monitoring. Kerma in air is the radiation field quantity used for the assessment of photon calibration fields with transfer ionisation chambers.

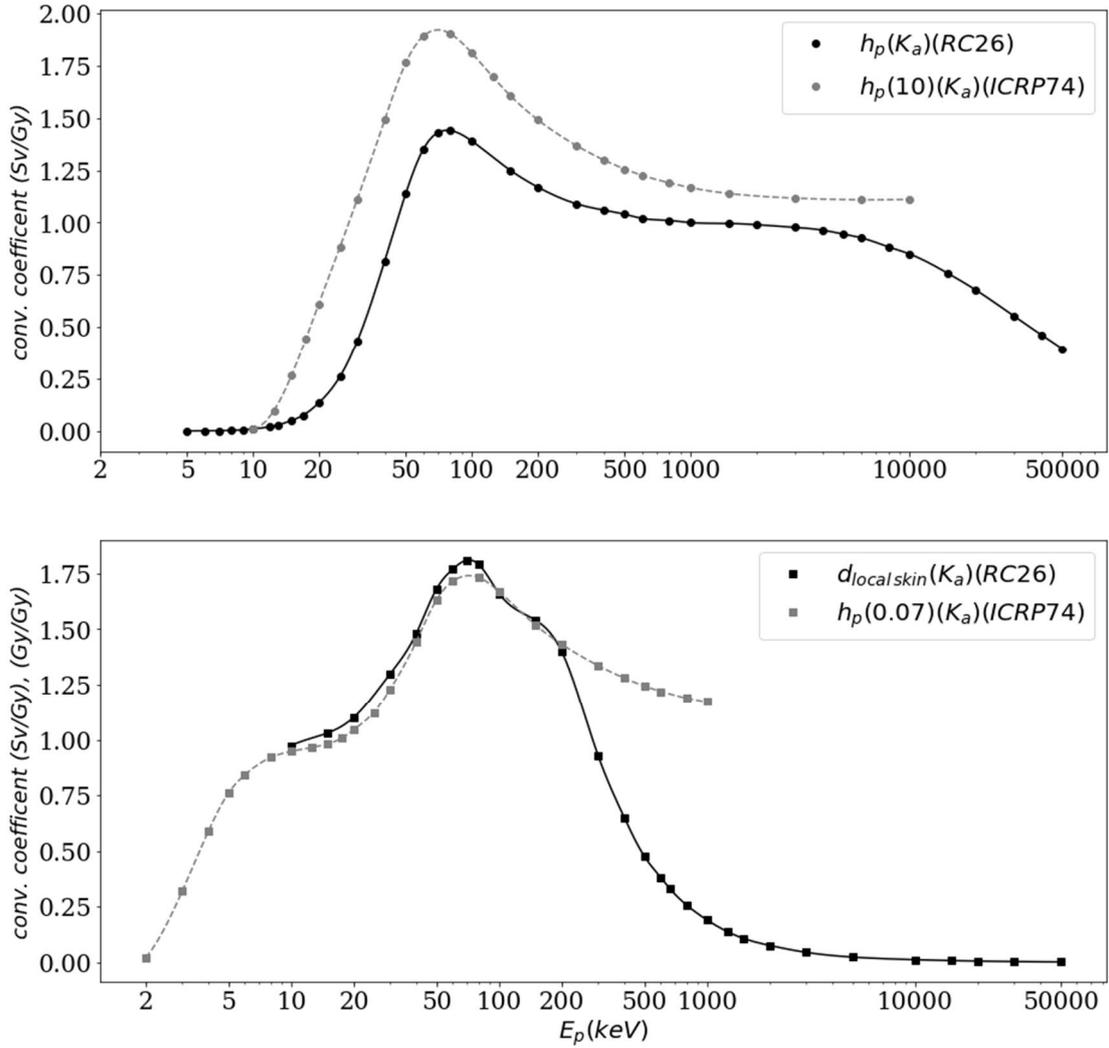

**Figure 2.** Monoenergetic conversion coefficients from kerma $K_a$ to operational quantities used for personal monitoring, irradiated anterior-posterior. (azimuthal angle $\alpha = 0^\circ$). Top: conversion coefficient from $K_a$ to operational quantities for whole-body monitoring: personal dose $H_p$ (black symbols) and personal dose equivalent $H_p(10)$ [2] (grey symbols). Bottom: conversion coefficient from $K_a$ to absorbed dose in local skin $D_{\text{local skin}}$ (black symbols) and to personal dose equivalent $H_p(0.07)$ (grey symbols).

Figure 2 shows in the top panel, that for all energies $E_p$ the presently used operational quantity $H_p(10, \alpha = 0^\circ)$ exceeds $H_p(\alpha = 0^\circ)$, which is numerically equivalent to effective dose $E$ (AP) in anterior-posterior irradiation geometry. At low energies, the overestimation is due to the assessment depth of 10 mm, which is not representative for the whole body with deep-seated organs, at high energy it is the consequence of the kerma-approximation used to simplify the radiation transport calculation. The lower panel shows, that $D_{\text{local skin}}$ is numerically very close to the presently used quantity $H_p(0.07)$ up to an energy $E_p \approx 200$ keV, from where on the kerma approximation introduces a difference.

## 2. Fluence-to-Dose Conversion Coefficients for X-ray spectra

Radiation protection dosimeters and survey instruments for photons are calibrated with reference radiations from X-ray generators and from radioisotope sources, described in standard ISO 4037-3 [6]. The Narrow-spectrum series is often employed for this purpose. Kerma-to dose conversion coefficients for these spectra are tabulated for the present operational quantities $H_p(10,\alpha)$ and $H^*(10)$ in [7], [8] and for $H_p(0.07,\alpha)$ in [9].

Generally, the conversion coefficient $c_k(S)$ from kerma to a quantity $C$ for a radiation quality $S$ with fluence spectrum $\phi_{E_p}(E_p)$ can be calculated by the expression

$$c_k(S) = \frac{\int \phi_{E_p}(E_p) \, k_\phi(E_p) \, c_k(E_p) dE_p}{\int \phi_{E_p}(E_p) \, k_\phi(E_p) dE_p} \, , \qquad (2)$$

where $c_k(E_p)$ are the monoenergetic conversion coefficients from kerma to quantity $C$. The integral is taken over the energy range covered by the fluence spectrum. The German National Standards laboratory PTB makes fluence spectra $\phi_{E_p}(E_p)$ and kerma-weighted fluence spectra $\phi_{E_p}(E_p) \, k_\phi(E_p)$ available on its website [10], with these data the conversion coefficients from kerma to $H^*$, $H_p$ and $D_{\text{local skin}}$ were calculated for the X-ray spectra of the Narrow-spectrum series according to Equation 2. Table 1 shows the result of this calculation for the incident angle $\alpha = 0°$.

**Table 1:** Spectrum-averaged conversion coefficients from kerma to the new operational quantities Ambient dose $h_k^*(S, 0°)$, Personal dose $h_{p,k}(S, 0°)$ and Absorbed dose to local skin $d_{\text{local skin}}(S, 0°)$. The average of the monoenergetic conversion coefficients for incidence under 0° (anterior-posterior geometry) extends over the spectra of the Narrow series from [6].

| Spectrum $S$ | $E_{\text{avg}}(S)$ (keV) | $h_k^*(S, 0°)$ (Sv/Gy) | $h_{p,k}(S, 0°)$ (Sv/Gy) | $d_{\text{local skin}}(S, 0°)$ (Gy/Gy) |
|---|---|---|---|---|
| N-10 | 8.5 | 3.94E-03 | 3.94E-03 | 0.927 |
| N-15 | 12.3 | 2.23E-02 | 2.23E-02 | 0.997 |
| N-20 | 16.3 | 6.16E-02 | 6.15E-02 | 1.039 |
| N-25 | 20.3 | 0.130 | 0.130 | 1.098 |
| N-30 | 24.6 | 0.235 | 0.235 | 1.178 |
| N-40 | 33 | 0.528 | 0.528 | 1.340 |
| N-60 | 48 | 1.026 | 1.026 | 1.607 |
| N-80 | 65 | 1.373 | 1.373 | 1.781 |
| N-100 | 83 | 1.424 | 1.424 | 1.755 |
| N-120 | 100 | 1.384 | 1.384 | 1.661 |
| N-150 | 118 | 1.327 | 1.327 | 1.599 |
| N-200 | 165 | 1.220 | 1.220 | 1.500 |
| N-250 | 207 | 1.160 | 1.160 | 1.344 |
| N-300 | 248 | 1.121 | 1.122 | 1.143 |
| S-Cs | 662 | 1.020 | 1.015 | 0.331 |
| S-Co | 1225 | 1.000 | 0.997 | |

## 3. Estimation of Dosimeter Response to New Operational Quantities

The aim of this section is to estimate the effect on instrument response of the possible introduction of the operational quantities recommended by ICRU RC 26 as legal quantities. Scientific publications and manufacturer's data were analyzed for of response functions of survey monitors and personal dosimeters. These are usually published as graphical representations, in which the response to one of the spectra from the ISO-4037-3 narrow series is marked with an entry at the average energy of the spectrum. Table 2 lists the survey instruments and dosimeters for which response functions were obtained, and their sources.

**Table 2:** Survey monitors and personal dosimeters for which response functions were obtained and analysed with respect to the newly proposed operational quantities.

| Dosimeter or Survey Instrument | Response Quantity | Source |
|---|---|---|
| Automess 6150 AD6 Geiger-Müller survey monitor | $H^*(10)$ | [11] |
| Centronics IG5 2 MPa Ar ionisation chamber | $H^*(10)$ | [12] |
| BEOSL environmental dosimeter | $H^*(10)$ | [13] |
| BEOSL personal dosimeter | $H_p(10), H_p(0.07)$ | [14] |
| HPA / PHE personal dosimeter | $H_p(10), H_p(0.07)$ | [15] |
| Mirion DMC 3000 electronic personal dosimeter | $H_p(10)$ | [16] |

The response of the dosimeter is noted as the ratio of the value indicated by the dosimeter $G$ over the conventional true value of the quantity $C$

$$R = \frac{G}{C}$$

If a change intervenes in the conventional true value $C$, for example by a new definition of the quantity, then the new response $R$ can be calculated from the known, old response $R_{old} = G/C_{old}$, via the old and new conversion coefficients, $c_{old}$ and $c$ respectively, by

$$R = \frac{G}{C_{old}} \frac{C_{old}}{C} = R_{old} \frac{C_{old}}{C} = R_{old} \frac{c_{old}}{c} \qquad (3)$$

This is possible if the method of calibration for the new and the old quantities is identical, as is the case with the operational quantities in use and newly defined. The conversion coefficients $c$ take the form of spectrum-averaged coefficients (equation 2), as listed in table 1 for the new operational quantities.

For example, the response of a survey monitor based on a Geiger-Müller counter is indicated as "counts per unit of ambient dose" or

$$R = \frac{N}{H^*}$$

When the response $R_{old}$ of the dosimeter to the presently used quantity $H^*(10)$ for a radiation spectrum $S$ is known, the response $R$ to the new quantity $H^*$ can be calculated as

$$R(S) = \frac{N}{H^*(10)} \frac{H^*(10)}{H^*} = R_{old}(S) \frac{H^*(10)}{H^*} = R_{old}(S) \frac{h^*(10,S)}{h^*(S)}$$

In the last step, the spectrum-averaged conversion coefficients are used. The relative response to the recommended quantity can be determined by normalizing to the response of the instrument under reference conditions. For photon dosimeters and area monitoring instruments, the reference condition is usually the radiation of the $^{137}$Cs isotope ($E_p$ = 662 keV), incident under 0°.

The plots of response functions from the sources in table 2 were digitized [17] and the obtained estimates of the spectral response to the presently used quantities $H^*(10)$, $H_p(10)$ and $H_p(0.07)$ were converted to response values to the new quantities $H^*$, $H_p$ and $D_{local\,skin}$ with help of equation (3).

## 4. Results

In the following graphs, the response functions of the analyzed instruments, retrieved from literature, are compared to the response functions to the new operational quantities, calculated according to equation (3) for each X-ray spectrum from the narrow series used in the original calibration. The data points are connected with smooth lines from a cubic spline interpolation, this serves to guide the eye.

The graphs show the relative response to the presently used quantities, normalized to the response at the reference energy $E_p$ = 662 keV. For the new quantities, the response is shown as if the instrument has not been recalibrated at the reference energy. Recalibrating the instruments would lower the response to the new quantities by approximately 20 % for H* and Hp.

### 4.1 Automess AD6150 AD6 Survey Monitor

The hand-held survey instrument Automess 6150 AD6 is based on a Geiger-Müller counter. The instrument has a sensitivity cut-off below $E_p$ = 50 keV and its energy response is thus insensitive to the strong variation of H* with respect to H*(10) at low energies (figure 3). Re-calibrating the survey instrument to H* at $E_p$ = 662 keV would make the two curves nearly indistinguishable, with a slightly better response to the N-60 spectrum for the new quantity.

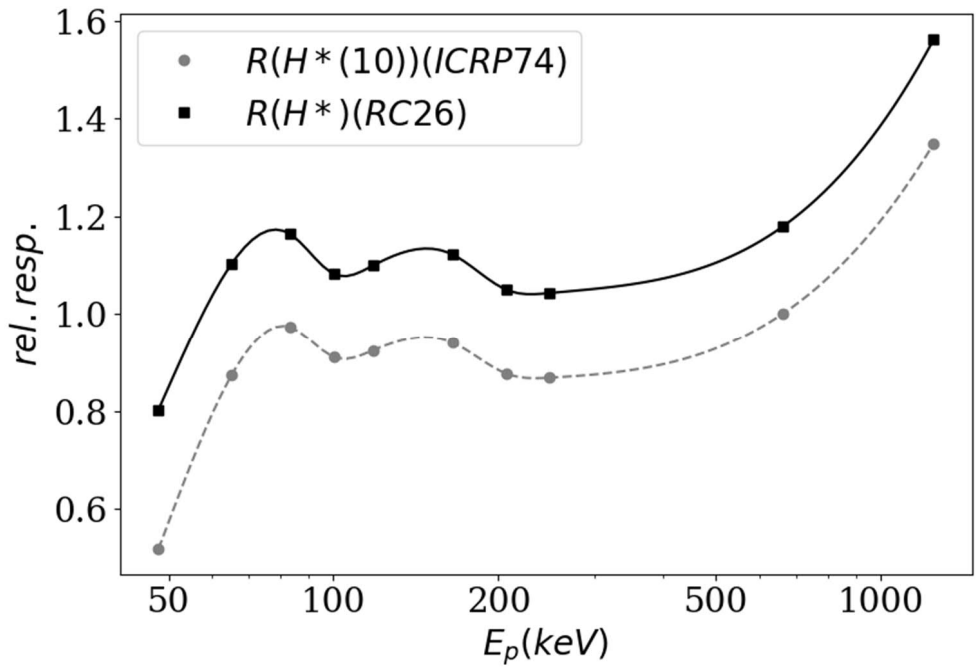

**Figure 3.** Response of the Automess 6150 AD6 survey meter [11] to ambient dose equivalent $H^*(10)$ (grey, dashed line) and to ambient dose $H^*$ (black, continuous line)

## 4.2 Centronics IG5 Ar 20 Ionisation Chamber

The stainless-steel walled ionization chamber IG5 Ar 20 is filled with Argon as at a pressure of p = 2 MPa. Like the G-M counter based survey monitor, it shows a rapid decrease of sensitivity below the N-80 spectrum (Figure 4). A re-calibration to H* at $E_p$ = 662 keV would make the two response curves virtually indistinguishable, showing that the instrument has the same measurement capability with respect to H* as to H*(10).

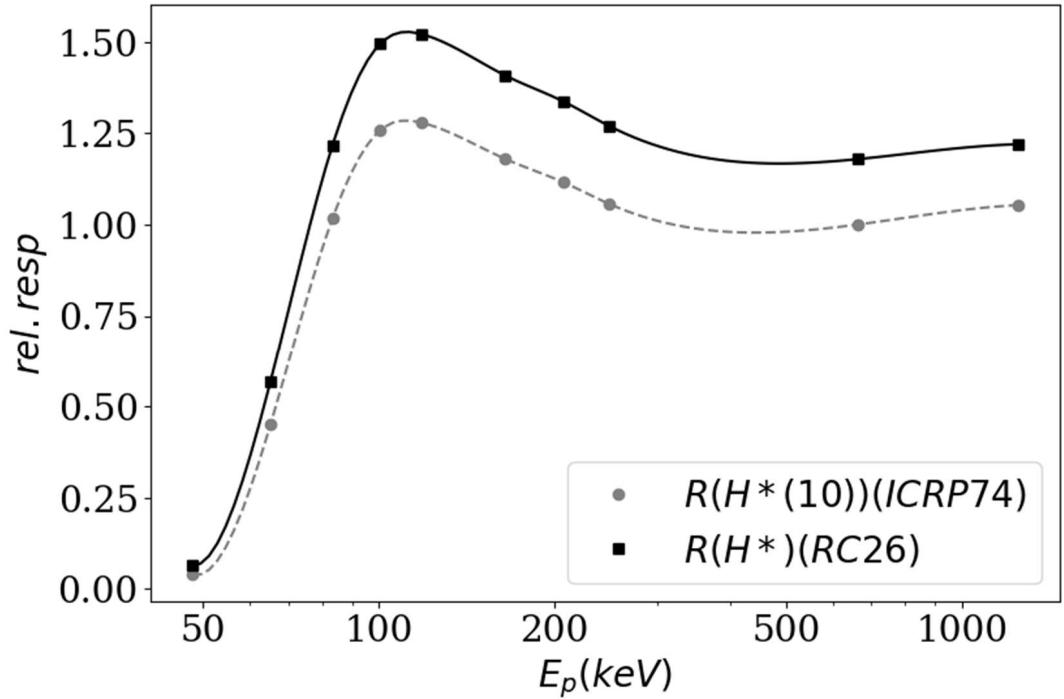

**Figure 4.** Response of the Centronics IG5 Ar 20 ionisation chamber [12] to ambient dose equivalent $H^*(10)$ (grey, dashed line) and to ambient dose $H^*$ (black, continuous line) and c.) personal dose equivalent $H_p(10)$ calculated with full electron transport (dashed line).

## 4.3 BEOSL Environmental Dosimeter

The BEOSL environmental dosimeter is based on a personal dosimeter (see below) with a modified holder to influence the energy response (Figure 5). Here we observe an overresponse of the dosimeter to $H^*$ at energies $E_p < 50$ keV, caused by the low-energy sensitivity required to model the energy dependence of $H^*(10)$.

## 4.4 BEOSL Personal Dosimeter

As the environmental dosimeter, the BeO-detector based personal dosimeter shows a similar overresponse to $H_p$ for low photon energies (Figure 6). The details of the energy response are determined by the dosimeter holder, optimized for a good response to $H_p(10)$. This feature will repeat for the other types of personal dosimeters.

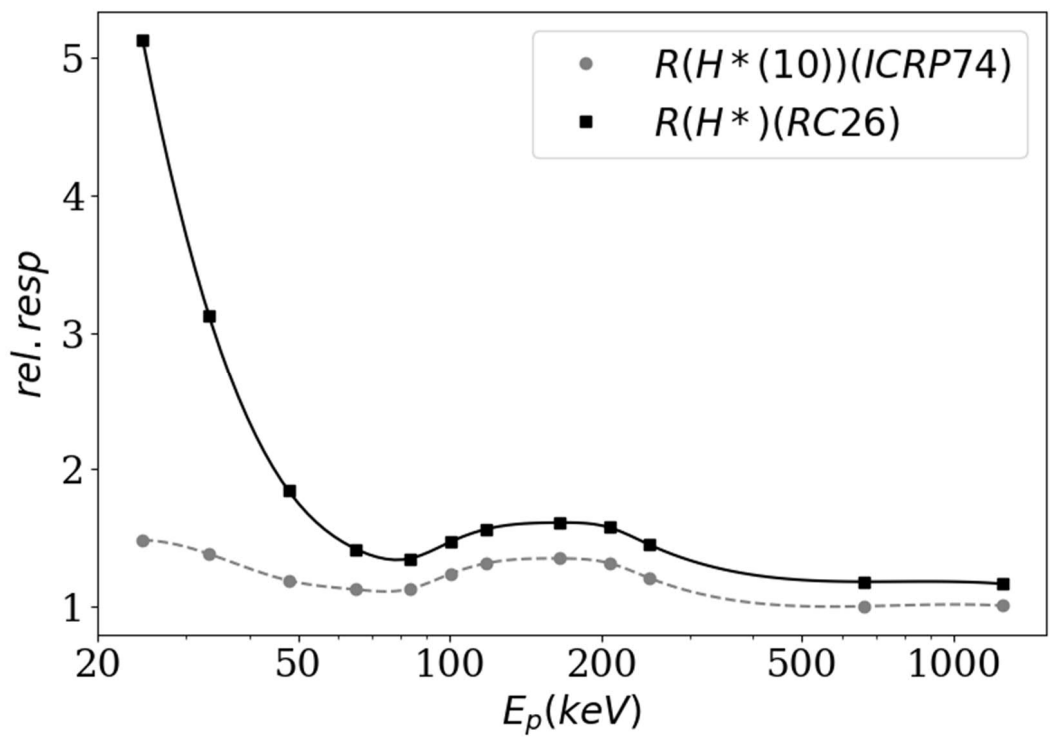

**Figure 5.** Response of the BEOSL passive environmental dosimeter [13] to ambient dose equivalent $H^*(10)$ (grey, dashed line) and to ambient dose $H^*$ (black, continuous line).

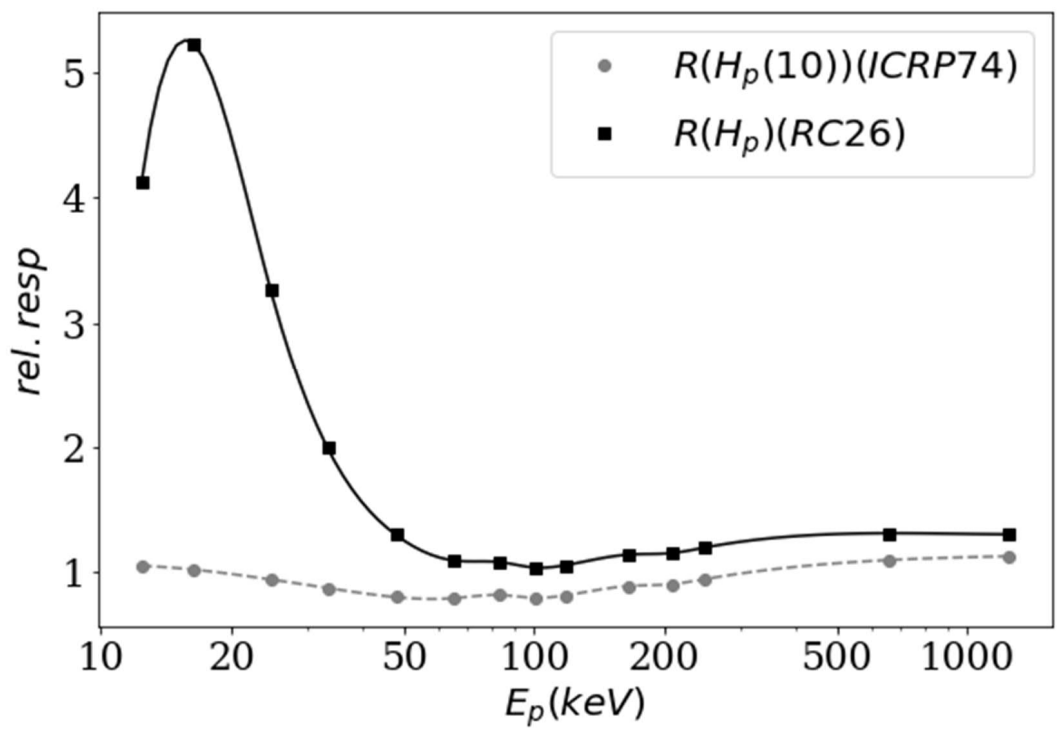

**Figure 6.** Response of the BEOSL passive personal dosimeter [14] to personal dose equivalent $H_p(10)$ (grey, dashed line) and to personal dose $H_p$ (black, continuous line).

**4.5 HPA / PHE Personal Dosimeter**

Public Health England (formerly Health Protection Agency) developed a passive dosimeter based on thermoluminescence detectors in a "Harshaw" card. The holder, conferring the energy response to the dosimeter, was first designed by Monte-Carlo simulations and then bult as a prototype and type-tested. As observed above, the good energy response to $H_p(10)$ leads to an overresponse to $H_p$ at photon energies $E_p < 50$ keV (Figure 7).

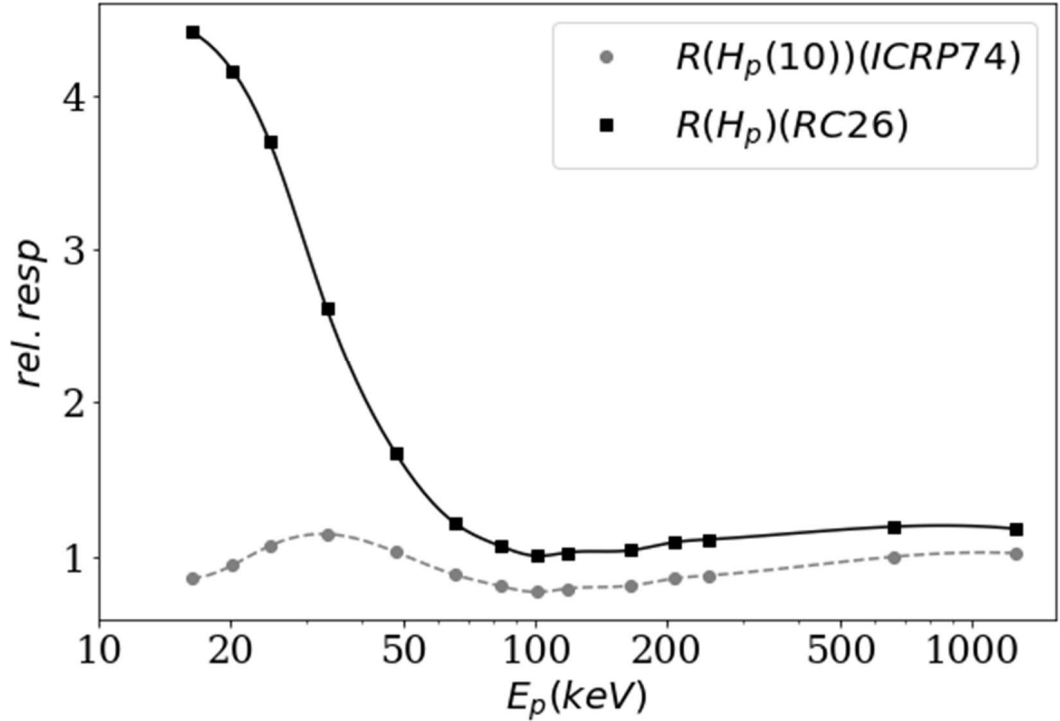

**Figure 7.** Response of the PHE / HPA passive personal dosimeter [15] to personal dose equivalent $H_p(10)$ (grey, dashed line) and to personal dose $H_p$ (black, continuous line).

**4.6 Mirion DMC 3000 Electronic Personal Dosimeter**

The DMC 3000 Personal Dosimeter is an electronic, direct reading dosimeter based on Si-diodes. In its basic version, the dosimeter measures only $H_p(10)$. The relative energy response to $H_p(10)$ and to $H_p$ are represented in Figure 8, showing the overresponse at low photon energies. At very low energies (Spectrum N-15), a sensitivity cut-off brings the response to $H^*$ back to an acceptable value.

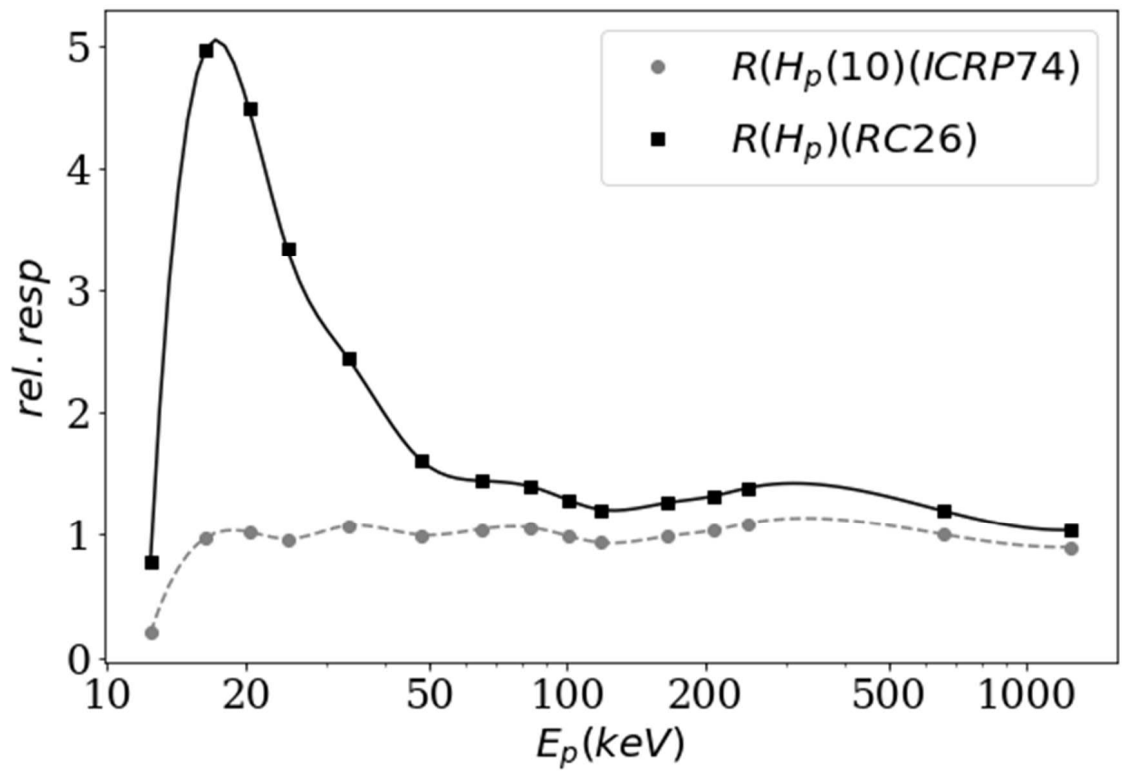

**Figure 8.** Response of the Mirion DMC 3000 electronic personal dosimeter [16] to personal dose equivalent $H_p(10)$ (grey, dashed line) and to personal dose $H_p$ (black, continuous line)

### 4.7 Response to the quantity $D_{local\,skin}$

The quantity *absorbed dose to local skin* $D_{local\,skin}$ is used for the assessment of deterministic radiation detriment, either to the extremities or to localized parts of the skin. In the proposal by ICRU RC 26 this quantity shall replace $H_p(0.07)$. In most countries, whole-body dosimeters must have the capability to measure $H_p(0.07)$. As can be seen from Figure 2, the difference of the two quantities is small for low energies up to $E_p \approx 200$ keV. Consequently, as shown in Figure 9, the two passive personal dosimeters for which relative response data to $H_p(0.07)$ were found in literature, perform well with respect to $D_{local\,skin}$ for all Narrow X-Ray spectra up to N-300. The deviation of the response at higher energies is irrelevant for personal dose monitoring because dose constraints for $H_p(10)$ or $H_p$ would be triggered before the much higher constraints for $H_p(0.07)$ or $D_{local\,skin}$ were crossed.

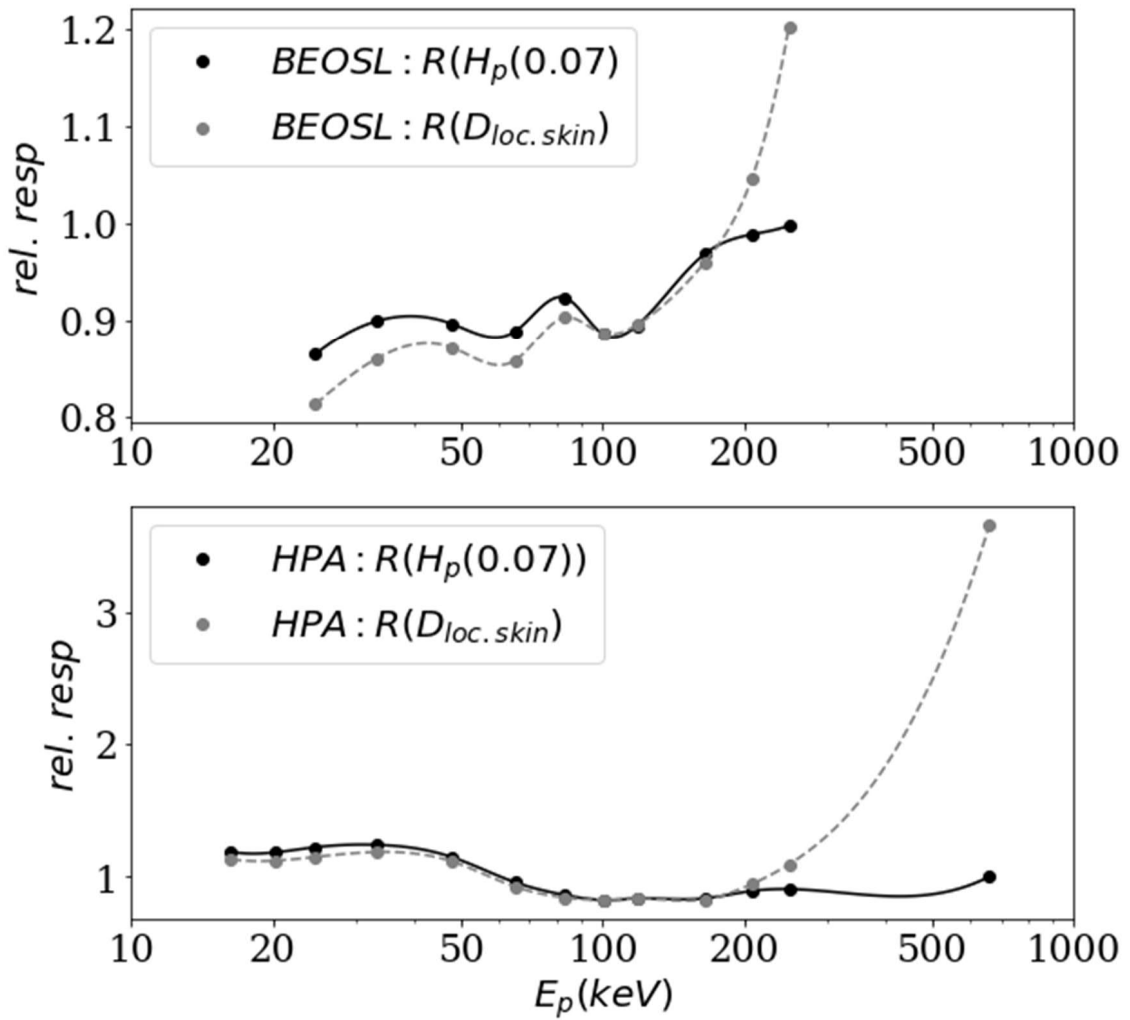

**Figure 9.** Top: Response of the BEOSL passive personal dosimeter [14] to to personal dose equivalent $H_p(0.07)$ (grey, dashed line) and to absorbed dose to local skin $D_{local\ skin.}$ (black, continuous line). Bottom: Response of the PHE / HPA passive personal dosimeter [15] to $H_p(0.07)$ (grey, dashed line) and to $D_{local\ skin.}$ (black, continuous line).

## 5. Discussion

To overcome the shortcomings of the presently introduced operational quantities for external radiation, ICRU Report Committee 26 proposes new operational quantities which are closely related to the protection quantities by using the same phantoms for the calculation of conversion coefficients. The purpose of this note was to evaluate the response of personal dosimeters and survey monitors to the new operational quantities.

From the limited range of investigated instruments in this note, three observations can be made:

1.) Survey instruments calibrated in ambient dose equivalent $H^*(10)$ can be recalibrated in ambient dose $H^*$ and then show very similar response characteristics, owing to the sensitivity cut-off for energies $E_p < 50$ keV. The response of the few available survey instruments which are sensitive to low-energy photons must be investigated separately.

2.) Personal dosimeters calibrated in personal dose equivalent $H_p(10)$ show an overresponse to $H_p$ at photon energies $E_p < 50$ keV. This is due to the fact that in this energy range $H_p(10) > E \approx H_p$.

3.) The detector elements in personal dosimeters for assessment of $H_p(0.07)$ can remain unmodified, they deliver good estimates of the new operational quantity $D_{\text{local skin}}$. The reason is that the definitions for the two quantities are very similar, they diverge only at high energies because of the use of the kerma approximation for $H_p(0.07)$. This energy region is without interest for monitoring $D_{\text{local skin}}$.

When the operational quantities proposed by RC 26 are introduced as legally binding for radiation protection measurements, the energy response of personal dosimeters must be adapted by changes to the dosimeter holder, or to the dose evaluation algorithm for multi-component dosimeters. The legislators must allow sufficient time for this process, to permit dosimeter designs to be developed which satisfy both technical and economical constraints.

## Acknowledgments

Discussions with the members of ICRU Report Committee are gratefully acknowledged: N. E. Hertel, D. T. Bartlett, R. Behrens, J.-M. Bordy, A. Endo, G. Gualdrini and M. Pellicioni.
This research did not receive any specific grant from funding agencies in the public, commercial, or not-for-profit sectors.